\documentclass[slac_one]{revtex4}
\usepackage{graphicx}
\usepackage{fancyhdr}
\begin{document}

\title{New and conventional bottomonium states}

\author{Jin Li}
\affiliation{Seoul National University, Seoul 151-747, Korea}

\begin{abstract}
Recent progress on the bottomonium states is reported.
This talk briefly reviews the observation of
$h_b(1P), h_b(2P), Z_{b1}^+, Z_{b2}^+$ states,
transition of $h_b(nP)\to \eta_b(mS)\gamma$
and new studies on the $\eta_b(2S)$ state.
Other $\eta$ and $\pi^+\pi^-$ transitions of
$\Upsilon(nS)$ are also discussed.
\end{abstract}

\maketitle

\thispagestyle{fancy}

\section{STUDY OF BOTTOMONIUM} 
The history of bottomonium starts in mid 1977 where an enhancement was
observed at a $\mu^+\mu^-$ mass peak near 9.5 GeV, from the collision
of 400 GeV protons on nuclear targets at Fermilab~\cite{Herb:1977ek}.
In 1978, with much improved resolution, the $\Upsilon$ particle was
observed with a width limited by energy spread of beams at the DORIS
$e^+e^-$ storage ring.  Just like the $J/\psi$ particle, the
$\Upsilon$ discovery comes from direct production, thanks for the
spin-parity $J^P=1^-$ and its substantial leptonic partial width.

\begin{figure}
\includegraphics[width=0.36\textwidth]{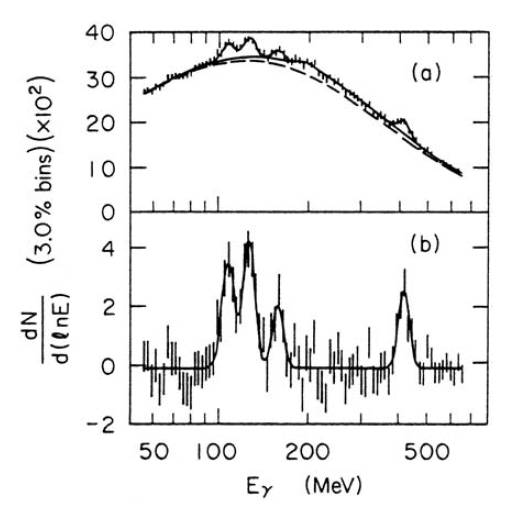}%
\includegraphics[width=0.64\textwidth]{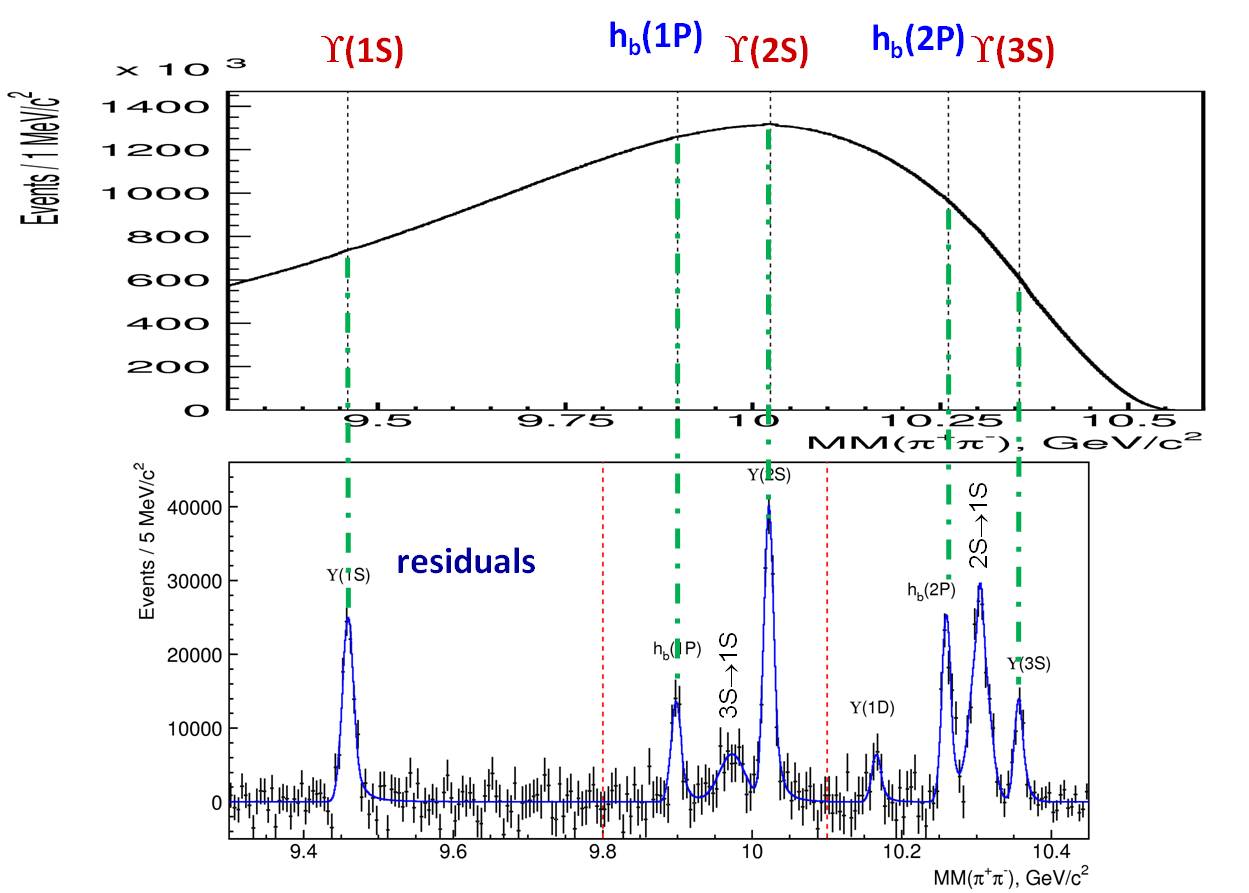}%
\caption{\label{fig:crystalball-h_b} Left figure shows the inclusive
photon spectrum obtained by the Crystal Ball Collaboration at DORIS
II~\cite{Nernst:1985nn}.  Right figure shows the inclusive missing
mass spectrum of $\pi^+\pi^-$ obtained at the Belle's 121.4 fb$^{-1}$
$\Upsilon(5S)$ data~\cite{Adachi:2011ji}, where the bottom plot is the
background subtracted residual distribution, with the fit function
overlaid.  Different dashed lines indicate the position of various
resonances.  Notice the similarity between left and right figures.}
\end{figure}

However, other particles of the bottomonium family can not be
directly produced at an $e^+e^-$ collider.  So, alternative
experimental methods should be used to search for these states.  For
example, the $P$-wave states of $b$ and $\bar b$ were discovered using
the inclusive photon spectrum of $\Upsilon'$ decay at Crystal Ball
collaboration in 1985.  In Fig.~\ref{fig:crystalball-h_b} left, A
triplet of peaks between 100 and at 200 MeV is clearly seen, which
corresponds to $\Upsilon'\to \gamma\chi_b(^3P_{2,1,0})$ transitions,
while the peak at the right is due to reflection from
$\chi_b\to\gamma\Upsilon$.

Now, a large of set bottomonium states has been predicted and studied,
which was nicely arranged according to the $b\bar b$ level scheme
(PDG).  Recent discoveries of new bottomonium states also follow the
history, where the inclusive or semi-inclusive methods play important
role.  As two examples not long ago, BaBar collaboration discovered
the lowest lying bottomonium state $\eta_b$ using the same inclusive
$\gamma$ spectrum method~\cite{Aubert:2008ba}, and CLEO collaboration
discovered $\Upsilon(1D)$ using the recoil mass of two soft $\gamma$s
with $l^+l^-$ tagging~\cite{Bonvicini:2004yj}, which represents the
inclusive property of two $\gamma$s.  In the following sections,
recent results of bottomonium, especially those obtained from Belle's
large 121.4 fb$^{-1}$ $\Upsilon(5S)$ data, are discussed.

\section{$\Upsilon(5S)$ and $h_b$ PARTICLES} 
In 2008, Belle collaboration reported surprising large partial widths
of $\Upsilon(5S)\to\Upsilon(1,2,3S)\pi^+\pi^-$~\cite{Abe:2007tk},
which are two orders of magnitude larger than the widths of
$\Upsilon(2,3,4S)\to\Upsilon(1S)\pi^+\pi^-$.  While the reason of this
discrepancy is unclear, an $\Upsilon$ state different from the
conventional $b\bar b$ bound state has been
considered~\cite{Chen:2008xia}.

In 2011, CLEO-c collaboration suggested that the cross-sections of
$h_c\pi^+\pi^-$ and $J/\psi\pi^+\pi^-$ are of similar magnitude, and
are all enhanced near 4.26 GeV $e^+e^-$ energy~\cite{CLEO:2011aa},
although for $h_c\pi^+\pi^-$ the statistics was limited.  Interpreting
this enhancement as the $Y(4260)$ resonance, the partial width of
$Y(4260)\to J/\psi \pi^+\pi^-$ has been shown to be greater than 508
keV at 90\% confidence level~\cite{Mo:2006ss}, which is much larger
than the partial widths of $\psi'$ (102 keV) and $\psi''$ (53
keV). This behavior is similar to the bottomonium case of
$\Upsilon(5S)$.  Thus, we naturally think that the $h_b\pi^+\pi^-$
production rate is also greatly enhanced in the $\Upsilon(5S)$ region,
by assuming the similar mechanism as in the charmonium case.

The $h_b$ decay modes are unknown and should be complicated, since it
will decay to light quarks.  In order to study the decay
$\Upsilon(5S)\to h_b\pi^+\pi^-$, we can still use inclusive method to
avoid reconstructing $h_b$ directly.  The recoil or missing mass of
$\pi^+\pi^-$ defined as
$MM(\pi^+\pi^-)=\sqrt{(P(\Upsilon(5S))-P(\pi^+\pi^-))^2} $ was studied
at Belle.  Here $P$ is the four momentum of relevant particles, with
the $P(\Upsilon(5S))$ obtained from beam energies.  The spectrum of
$MM(\pi^+\pi^-)$ shown in Fig.~\ref{fig:crystalball-h_b} right is of
huge statistics which is around 1 million events per 1 MeV.
Nevertheless, it shows the similar behavior compared to the Crystal
Ball's inclusive photon plot in Fig.~\ref{fig:crystalball-h_b} left.  By
fitting it, Belle made the first observation of $h_b(1P)$ and
$h_b(2P)$ states~\cite{Adachi:2011ji}, whose masses agree with the
theoretical expectation from center-of-gravity of $\chi_b$ states.
However, the mechanism of $\Upsilon(5S)\to h_b(nP)\pi^+\pi^-$ decay is
exotic, because the ratio of spin-flip to non-spin-flip branching
fractions $\Gamma(\Upsilon(5S)\to h_b(nP)\pi^+\pi^-)/
\Gamma(\Upsilon(5S)\to \Upsilon(2S)\pi^+\pi^-) $ are not suppressed.

\section{$Z_b$ PARTICLES}
\begin{figure}
\parbox[c]{0.33\textwidth}
{\includegraphics[width=0.33\textwidth]{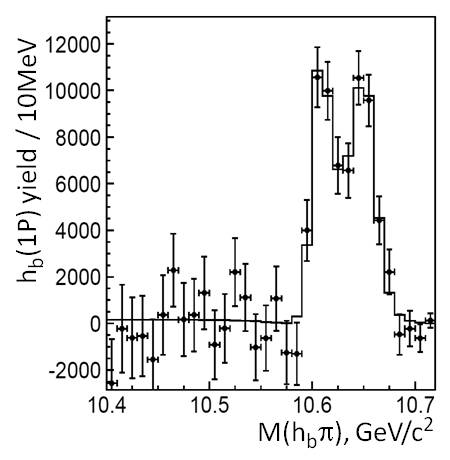}}%
\hspace*{0.1\textwidth}%
\parbox[c]{0.33\textwidth}
{\includegraphics[width=0.33\textwidth]{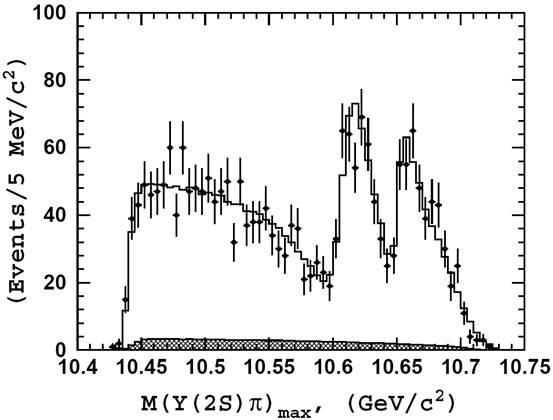}}%
\caption{\label{fig:Z_b}
The $M(h_b(1P)\pi)$ (left) and $M[\Upsilon(2S)\pi]_\mathrm{max}$ (right)
distributions in the $Z_b$ observation study at Belle~\cite{Belle:2011aa}.
Two peaks in both plots show the $Z_{b1}$ and $Z_{b2}$ signals.}
\end{figure}

While CLEO-c did not report any resonance structure in the
$h_c\pi^+\pi^-$ system, Belle has enough statistics to study the
resonance structure of the three body $h_b\pi^+\pi^-$ in
$\Upsilon(5S)$ decay.  This was achieved by looking the missing mass
of a single $\pi^+$ or $\pi^-$, which should effectively be the
$M(h_b\pi^-)$ or $M(h_b\pi^+)$.  Because of symmetry transposing
$\pi^+$ and $\pi^-$, the two missing mass distributions is combined
and upper half of the available range is used, which we denote as
$MM(\pi)$ distribution.  Then the missing mass of two pions
$MM(\pi^+\pi^-)$ was fitted to extract $h_b(1P)$ and $h_b(2P)$ signals
in bins of $MM(\pi)$.  The resulting spectra of $h_b$ yields as a
function $MM(\pi)$ will be background-free distributions for
$M(h_b(1P)\pi)$ and $M(h_b(2P)\pi)$.  The distribution for
$M(h_b(1P)\pi)$ from data exhibits a clear two-peak structure without
significant non-resonance component (Fig.~\ref{fig:Z_b} left).  The
distribution $M(h_b(2P)\pi)$ behaves similarly with smaller
statistics.  These two structures are referred as $Z_{b1}$ and
$Z_{b2}$ and parameterized as two $P$-wave Breit-Wigner amplitudes.
The fit function with $\sqrt{s}=M(h_b\pi)$ is:
\begin{equation}
|\mathrm{BW}(s,M_1,\Gamma_1)+ae^{i\phi}\mathrm{BW}(s,M_2,\Gamma_2)
+ be^{i\psi}|^2\frac{qp}{\sqrt{s}}.
\end{equation}
Here the amplitude of two resonances and a non-resonant component
are added coherently to form the rate, which is then multiplied with
a phase-space factor $\frac{qp}{\sqrt{s}}$, where $q$ or $p$ is
the momentum of the pion from $\Upsilon(5S)$ or $Z_b$ decay in the
rest frame of their mother particles.

The $Z_b$ particles can also be studied from the decay
$\Upsilon(5S)\to\Upsilon(nS)\pi^+\pi^-$, with $n=1,2,3$.  In this
case, the final state particles can be fully reconstructed with
$\Upsilon(nS)\to\mu^+\mu^-$.  Dalitz analysis is then performed to the
three-body final states to extract maximal information.  The
amplitudes used in the parametrization includes two Breit-Wigners for
$Z_{b1}$ and $Z_{b2}$, a coupled-channel Breit-Wigner for $f_0(980)$
scalar, a Breit-Wigner for $f_2(1270)$ tensor state, and a
non-resonant amplitude.  Results of $Z_b$ parameters for all five
channels are consistent, and the average masses and widths are
$M=10607.2\pm2.0$ MeV, $\Gamma=18.4\pm2.4$ MeV for $Z_{b1}$ and
$M=10652.2\pm1.5$ MeV, $\Gamma=11.5\pm2.2$ MeV for
$Z_{b2}$~\cite{Belle:2011aa}.

If we compare the relative phase of two $Z_b$s in $h_b(1P)\pi^+\pi^-$
channel and $\Upsilon(2S)\pi^+\pi^-$ channel, we found that the former
is close to 180$^\circ$ and the latter is close to zero degree.  From
the distribution of $M(\Upsilon(2S)\pi)_\mathrm{max}$, as shown in
Fig.~\ref{fig:Z_b} right, a large destructive interference dip near
10.6 GeV/$c^2$ suggests that this interference effect exists in the
whole Dalitz plane, which is only possible if the $Z_b$ has a
spin-parity $J^P=1^+$.

\section{$\eta_b$ PARTICLES}
$h_b$ is expected to have large decay branching fractions to
$\eta_b\gamma$~\cite{Godfrey:2002rp}.  The $\eta_b$ particle
previously has been studied~\cite{Aubert:2008ba,Bonvicini:2004yj} at
BaBar and CLEO in the channel $\Upsilon(3S)\to\eta_b\gamma$.  Since
large $h_b$ samples are available at Belle, it is natural to study the
transition from $h_b$ to $\eta_b\gamma$.

\begin{figure}
\setlength{\unitlength}{0.005\textwidth}
\begin{tabular}{c}
\includegraphics[width=0.5\textwidth]{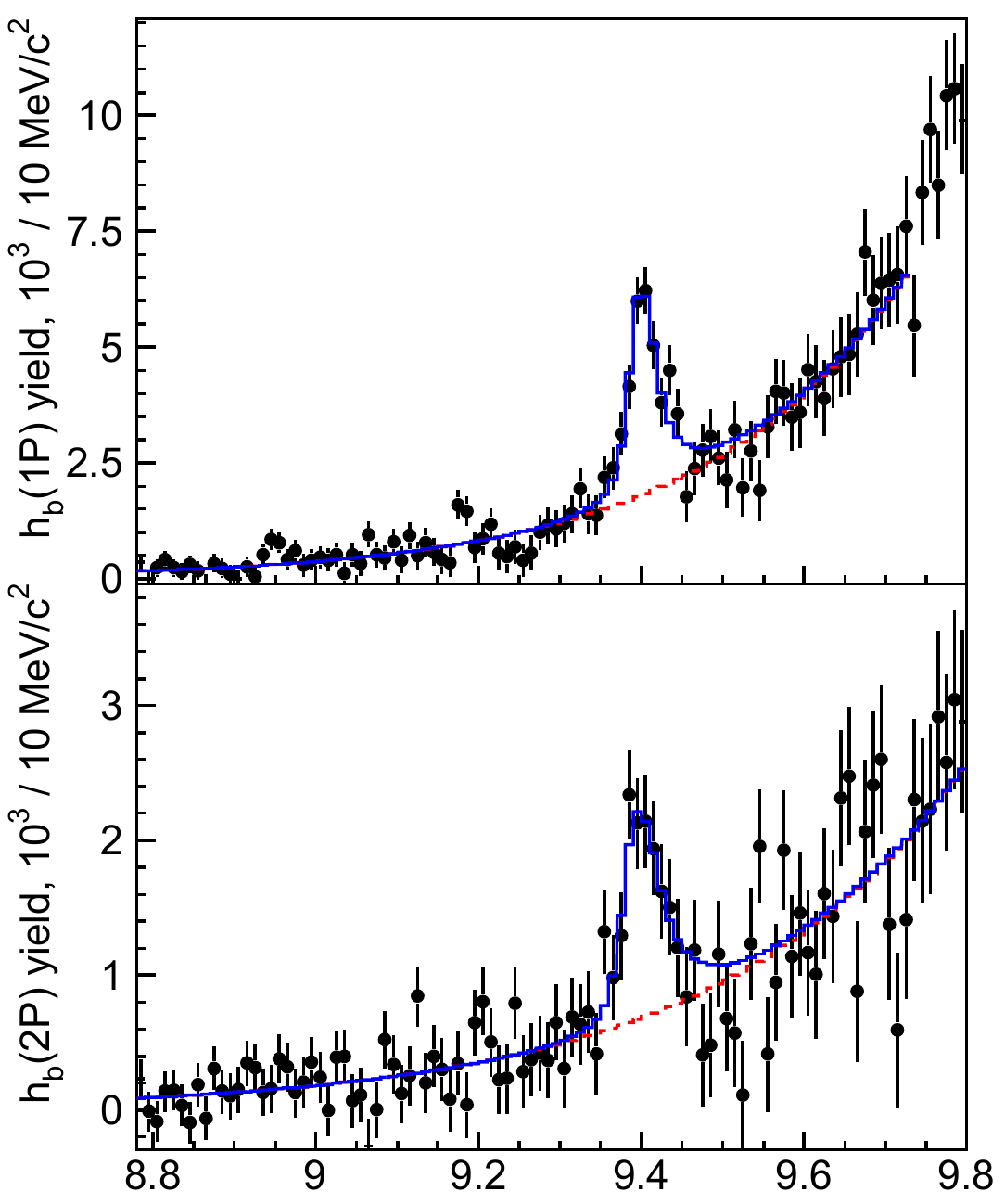}%
\begin{picture}(0,0)
\put(-75,110){(a)}
\put(-75,53){(b)}
\end{picture}\\
$\Delta MM(\pi^+\pi^-\gamma)$, GeV/$c^2$
\end{tabular}%
\begin{tabular*}{0.5\textwidth}{c}
\includegraphics[width=0.5\textwidth]{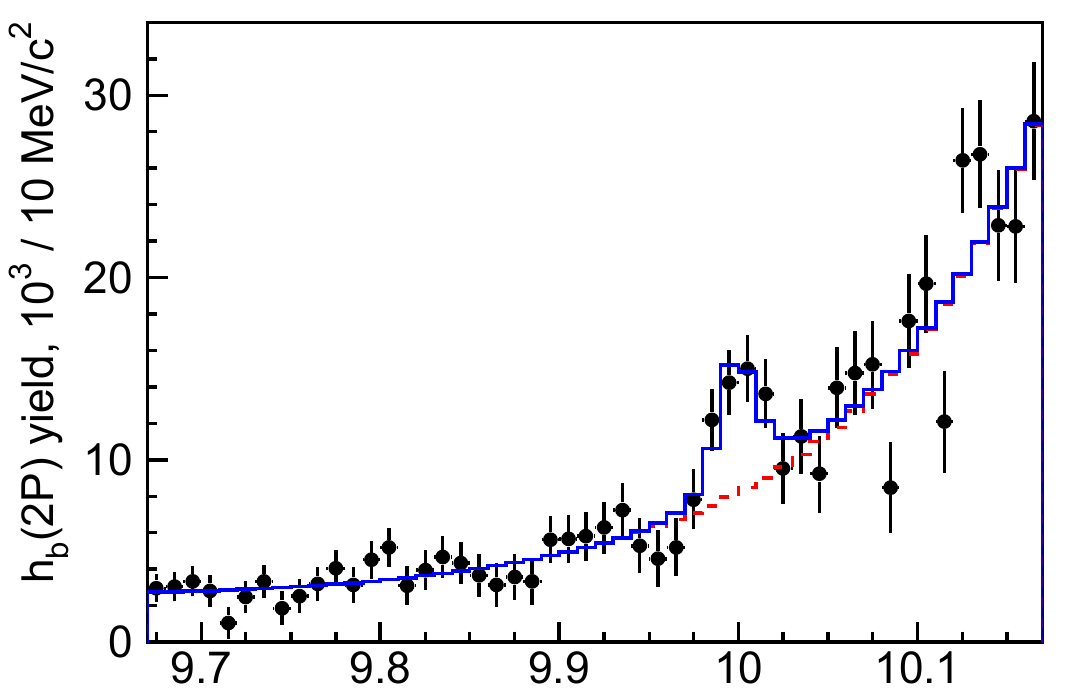}%
\begin{picture}(0,0)\put(-75,53){(c)}\end{picture}\\
$\Delta MM(\pi^+\pi^-\gamma)$, GeV/$c^2$\\
\includegraphics[width=0.5\textwidth]{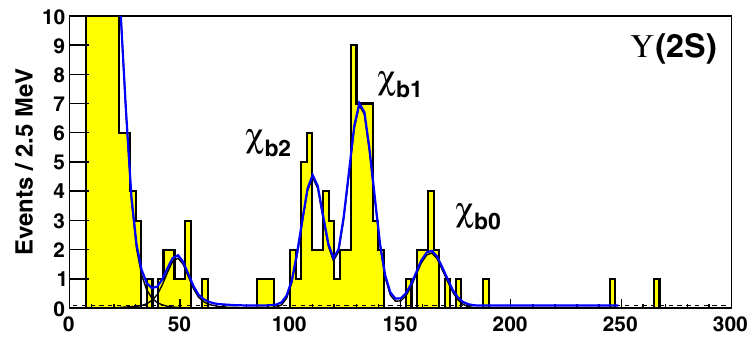}%
\begin{picture}(0,0)\put(-75,35){(d)}\end{picture}\\
$\Delta M$ (MeV/$c^2$)\\
\end{tabular*}%
\caption{\label{fig:etab-1S2S} For Belle data~\cite{Mizuk:2012pb}, the
$h_b(1P)$, $h_b(2P)$ yields in the $\eta_b(1S)$ region are shown
in (a), (b) and the $h_b(2P)$ yield in the $\eta_b(2S)$ region is
shown in (c).  Clear $\eta_b(1S)$ and $\eta_b(2S)$ signals are all
seen in (a,b,c).  The mass difference between $\Upsilon(2S)$ and
$\eta_b(2S)$ candidates in CLEO data~\cite{Dobbs:2012zn} is shown in
(d), where the unidentified small peak around 50 MeV is interpreted as
the $\eta_b(2S)$ signal.  }
\end{figure}
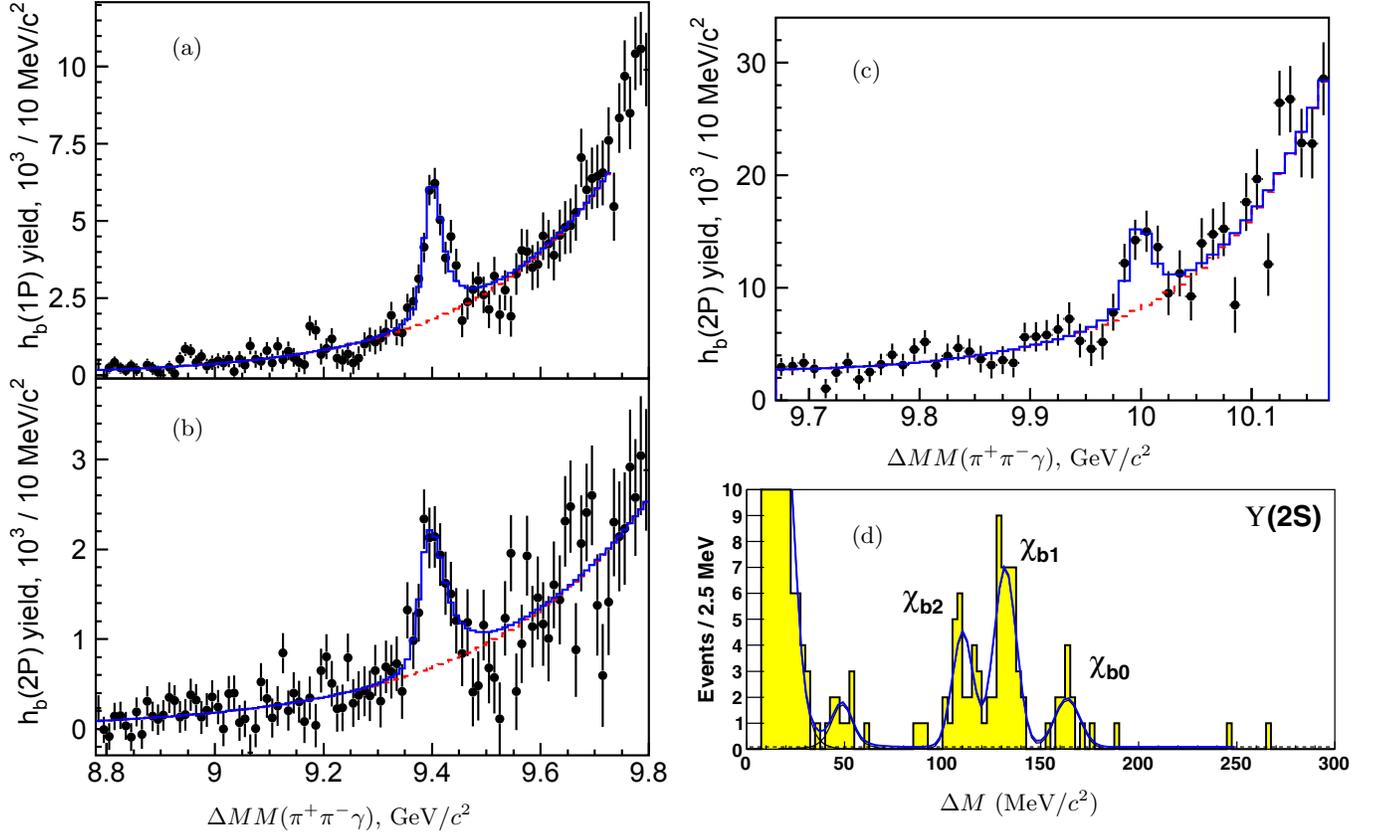

The method to extract $\eta_b$ signals at Belle is similar to what is
used in extracting $h_b$ signals.  Here the missing mass of three
particles $MM(\pi^+\pi^-\gamma)$ is calculated.  To avoid the
correlation between $MM(\pi^+\pi^-\gamma)$ and $MM(\pi^+\pi^-)$, their
difference is used, as $\Delta MM(\pi^+\pi^-\gamma) =
MM(\pi^+\pi^-\gamma)-MM(\pi^+\pi^-)+M(h_b)$, where $M(h_b)$ is fixed
at the nominal value.  The $MM(\pi^+\pi^-)$ spectra are then fitted in
different $\Delta MM(\pi^+\pi^-\gamma)$ bins, to obtain the
$h_b(1P,2P)$ yields as a function of $\Delta MM(\pi^+\pi^-\gamma)$, as
shown in Fig.~\ref{fig:etab-1S2S} (a,b,c).  The fact that $h_b$ is
produced from the $Z_b$ resonances is used to suppress the background
by the requirement $10.59
\mathrm{GeV}/c^2<MM(\pi)<10.67\mathrm{GeV}/c^2$.

The $h_b(1P,2P)$ yields as a function of $\Delta MM(\pi^+\pi^-\gamma)$
in $\eta_b(1S)$ and $\eta_b(2S)$ mass regions are fitted to a sum of
the $\eta_b$ signal component and a smooth background component, for
the transitions $h_b(1P)\to\eta_b(1S)\gamma$,
$h_b(2P)\to\eta_b(1S)\gamma$, and $h_b(2P)\to\eta_b(2S)\gamma$.  The
signal is described by the convolution of a non-relativistic
Breit-Wigner function with the resolution.  At the time of this
report, the fitted masses and widths are
$m_{\eta_b(1S)}=9402.4\pm1.5\pm1.8\;\mathrm{MeV}/c^2$,
$\Gamma_{\eta_b(1S)}=10.8^{+4.0}_{-3.7}{}^{+4.5}_{-2.0}\;\mathrm{MeV}$,
and $m_{\eta_b(2S)}=9999.0\pm3.5^{+2.8}_{-1.9}\;\mathrm{MeV}/c^2$%
~\cite{Mizuk:2012pb}.  The hyperfine splittings
$m_{\Upsilon(nS)}-m_{\eta_b(nS)} $ are determined as $\Delta
M_{HF}(1S) = 57.9\pm 2.3\;\mathrm{MeV}/c^2$, $\Delta M_{HF}(2S) =
24.3^{+4.0}_{-4.5}\;\mathrm{MeV}/c^2$.  This agrees with the
theoretical calculations~\cite{Meinel:2010pv}.  S.Dobbs \textit{et
al.} studied the $\eta_b(2S)$ in the process $\Upsilon(2S)\to
\gamma\eta_b(2S)$ in CLEO data, by reconstructing 26 exclusive modes
together for the $\eta_b(2S)$ candidate.  The hyperfine splitting is
reported as $\Delta M_{HF}(2S) = 48.7\pm
2.7\;\mathrm{MeV}/c^2$~\cite{Dobbs:2012zn}.  Fig.~\ref{fig:etab-1S2S}
(d) shows the fit.  This is of about 5 sigma discrepancy compared to the
Belle result and is in strong disagreement with theory.  So further
experimental clarification is needed.

\section{$\eta$ TRANSITIONS of $\Upsilon(nS)$
and other $\Upsilon(5S)$ decays}
The transition $\Upsilon(nS)\to \eta\Upsilon(mS)$ is a spin-flip E1M2
transition.  From the QCD multipole formalism~\cite{Voloshin:2007dx},
this spin-slip amplitude scales as $1/m_b$, and this $\eta$ transition
is suppressed compared to the $\pi\pi$ transition.  However, the
experimental values do not support those predictions.  Scaling from
the known branching fraction $\psi'\to\eta J/\psi$, the branching
fraction of $\Upsilon(2S)\to\Upsilon(1S)\eta$ should be around
$8\times 10^{-4}$, but the experimental value is around $2\times
10^{-4}$~\cite{He:2008xk}.  In addition, from
Ref.~\cite{Aubert:2008az}, the relation
$\mathcal{B}(\Upsilon(4S)\to\Upsilon(1S)\eta)\simeq 2.5\times
\mathcal{B}(\Upsilon(4S)\to\Upsilon(1S)\pi^+\pi^-)$ contradicts with
the expected suppression in $\eta$ transition.  So it is important to
study more in these channels.

\begin{figure}
\includegraphics[width=0.33\textwidth]{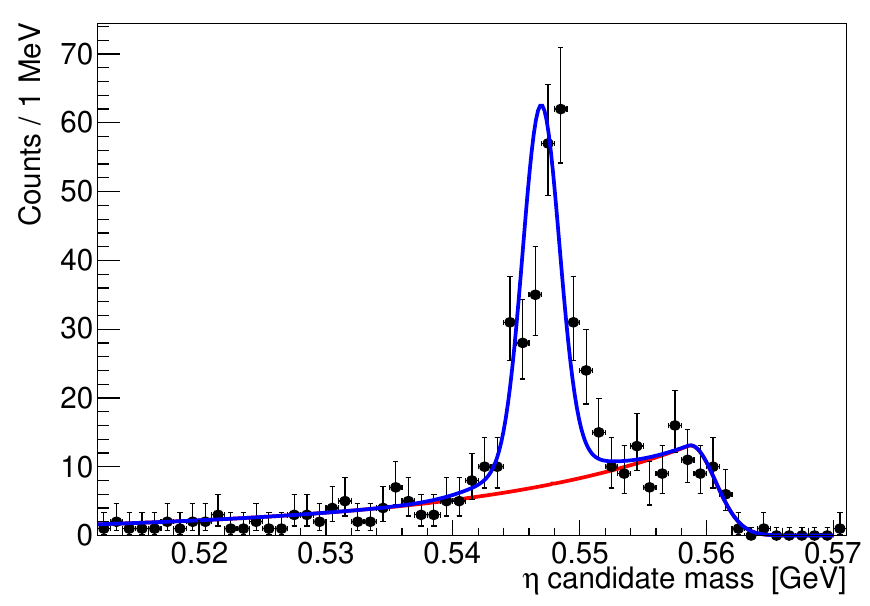}%
\includegraphics[width=0.33\textwidth]{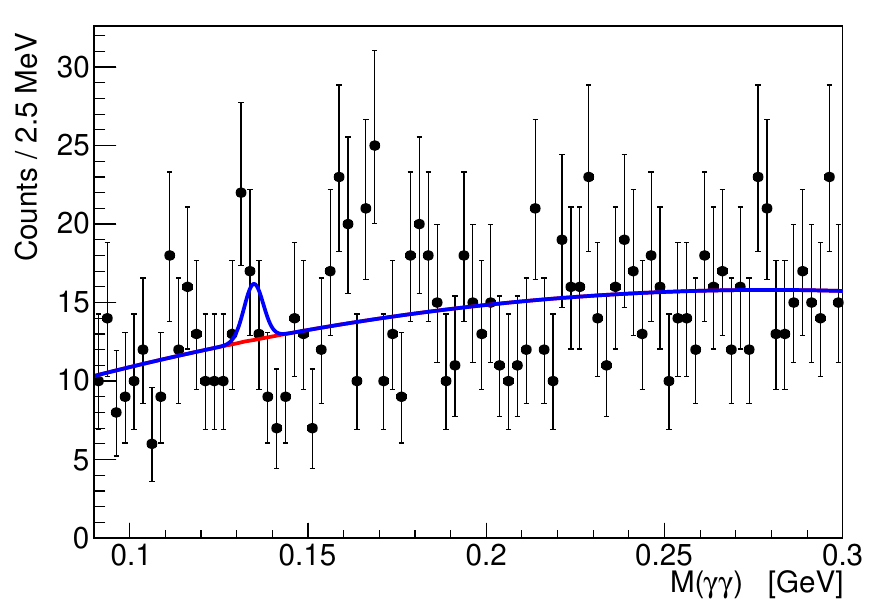}%
\caption{\label{fig:Y1S_eta}
$\eta$ (left) and $\gamma\gamma$ (right) mass distributions in the
$\Upsilon(2S)\to\eta\Upsilon(1S)$ and $\Upsilon(2S)\to\pi^0\Upsilon(1S)$ 
search at Belle.  $\eta$ candidates from $\gamma\gamma$ and $\pi^+\pi^-\pi^0$
channels are combined.}
\end{figure}

Belle performed a study of $\Upsilon(2S)\to(\eta,\pi^0)\Upsilon(1S)$
using Belle's 24.7 fb$^-1$ $\Upsilon(2S)$ data.  The reconstruction of
$\Upsilon(1S)\to e^+e^-,\mu^+\mu^-$ and a small total reconstructed
momentum in the $\Upsilon(5S)$ center-of-mass frame are required.  By
fitting the $\eta$ candidate and $\gamma\gamma$ masses, as shown in
Fig.~\ref{fig:Y1S_eta} we obtain
$\mathcal{B}(\Upsilon(2S)\to\Upsilon(1S)\eta)=
(3.41\pm0.30\pm0.35)\times 10^{-4}$ and
$\mathcal{B}(\Upsilon(2S)\to\Upsilon(1S)\pi^0)< 4.3\times 10^{-5}$
(90\% CL).

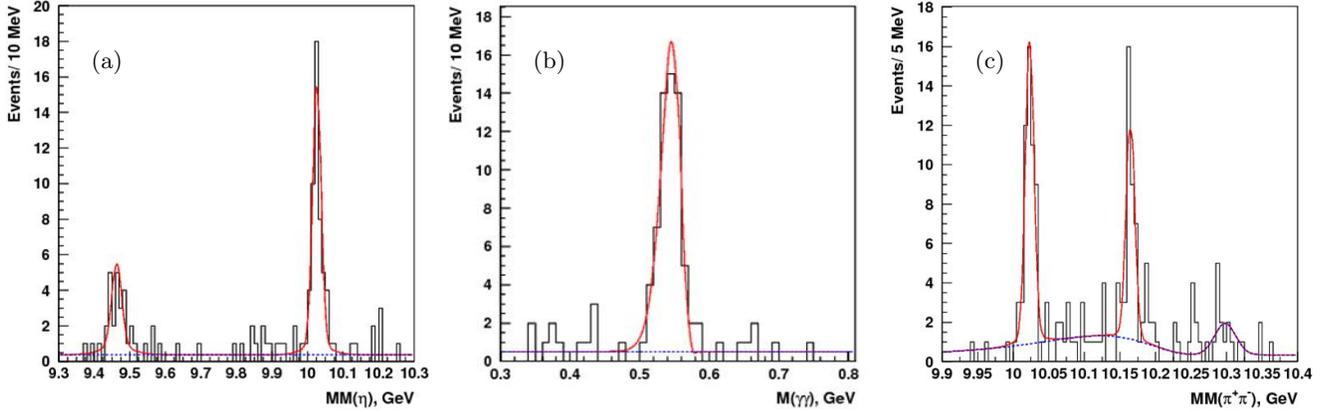
\begin{figure}
\setlength{\unitlength}{0.005\textwidth}
\includegraphics[width=0.33\linewidth]{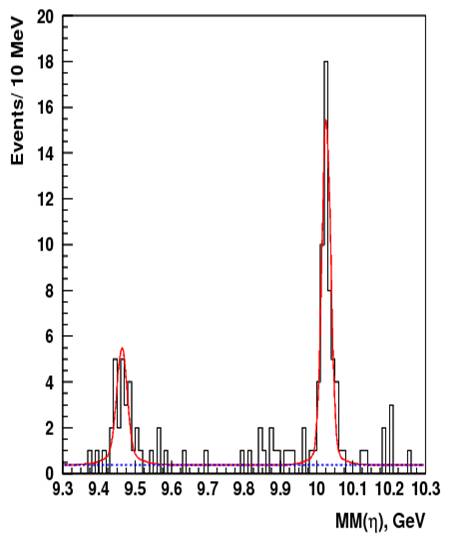}%
\begin{picture}(0,0)\put(-52,52){(a)}\end{picture}%
\includegraphics[width=0.33\linewidth]{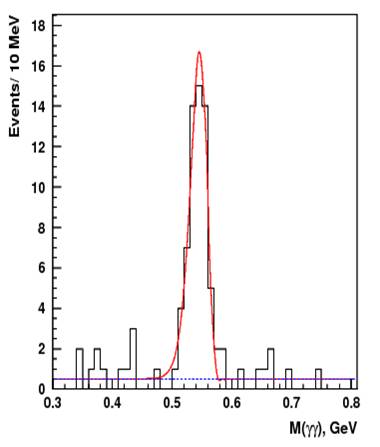}%
\begin{picture}(0,0)\put(-52,52){(b)}\end{picture}%
\includegraphics[width=0.33\linewidth]{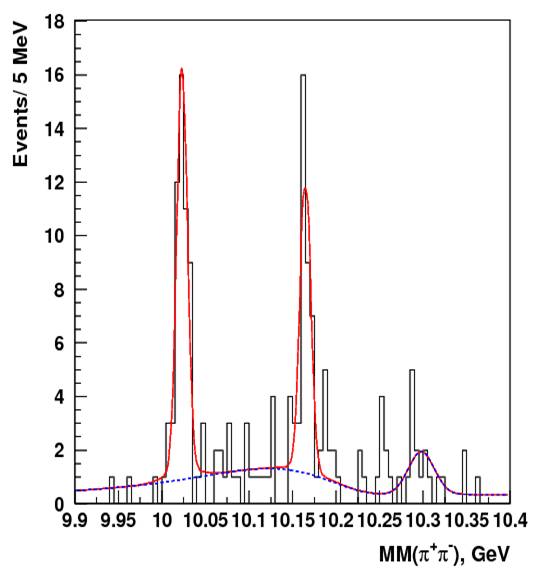}%
\begin{picture}(0,0)\put(-52,52){(c)}\end{picture}%
\caption{\label{fig:pipigg}
Plots of $\Upsilon(5S)\to\Upsilon(1,2S)\pi^+\pi^-\gamma\gamma$
study at Belle.
Look-back plots for the missing mass of $\eta$ in
the $\Upsilon(5S)\to\Upsilon(1,2S)\eta(\pi^+\pi^-\pi^0)$
study (a) and the $\gamma\gamma$ mass in
the $\Upsilon(5S)\to\Upsilon(2S)(\Upsilon(1S)\pi^+\pi^-)\eta$ study (b)
are shown. Missing mass of $\pi^+\pi^-$ after the
$\chi_{b1,2}\to\Upsilon(1S)\gamma$ selection is shown in (c), where
the left, middle, right peaks are due to $\Upsilon(2S)\pi^+\pi^-$,
the signal $\Upsilon(1D)\pi^+\pi^-$, and the
$\Upsilon(2S)(\Upsilon(1S)\pi^+\pi^-)\eta(\gamma\gamma)$ reflection.
}
\end{figure}

The $\Upsilon(5S)$ data was also analyzed similarly to search for
transitions to $\Upsilon(1,2S)\pi^+\pi^-\gamma\gamma$ states,
with $\Upsilon(1,2S)$ reconstructed in the $\mu^+\mu^-$ channel.
If we require the $\eta\to\pi^+\pi^-\pi^0$ selection criterion
for the pions and photons, $\eta$ transition of $\Upsilon(5S)$
can be studied. At the same time, if we require the
$\Upsilon(2S)\to\Upsilon(1S)\pi^+\pi^-$ selection,
$\Upsilon(5S)\to\Upsilon(2S)\eta$ can also be studied, since the
remaining two $\gamma$s can make an $\eta$. This time
we simply fit the difference of the $\Upsilon(5S)$ candidate's energy and
beam energy in the center-of-mass system to extract the signal.
This method gave the same result as fitting the missing mass of
$\eta$.  We found $\mathcal{B}(\Upsilon(5S)\to\Upsilon(1S)\eta)=
(7.3\pm1.6\pm0.8)\times 10^{-4}$, 
$\mathcal{B}(\Upsilon(5S)\to\Upsilon(2S)\eta)=
(38\pm4\pm5)\times 10^{-4}$, and
$\mathcal{B}(\Upsilon(5S)\to\Upsilon(1S)\eta')<
1.2\times 10^{-4}$ (90\% CL).  Fig.~\ref{fig:pipigg} (a,b) shows the
look-back plots of the first two modes.

A peak of $\Upsilon(1D)$ was seen in the inclusive distribution of
missing mass of $\pi^+\pi^-$ where the $h_b$ states was observed
(Fig.~\ref{fig:crystalball-h_b}).  A more complete study of the decay
$\Upsilon(5S)\to\Upsilon(1D)\pi^+\pi^-$ is now possible when the
exclusive mode $\Upsilon(1,2S)\pi^+\pi^-\gamma\gamma$ is being
analyzed, since we can aim for the sequential decay of
$\Upsilon(1D)\to\chi_b\gamma\to\Upsilon(1S)\gamma\gamma$.  The peak of
$\Upsilon(1D)$ became clearer after this exclusive reconstruction, and
turned out to be even sharper after the
$\chi_{b1,2}\to\Upsilon(1S)\gamma$ selection.  From the fit shown in
Fig.~\ref{fig:pipigg} (c), we obtain the observation of this channel
with $\mathcal{B}(\Upsilon(5S)\to\Upsilon(1D) \pi^+\pi^-)\times
\mathcal{B}(\Upsilon(1D)\to\chi_{b1,2}(1P)\gamma)\times
\mathcal{B}(\chi_{b1,2}(1P)\to\Upsilon(1S)\gamma) =
(2.0\pm0.4\pm0.3)\times 10^{-4}$.

\section{SUMMARY}
Belle's large $\Upsilon(5S)$ data provided many new results of
bottomonium states. The inclusive spectra of $\gamma, \pi^0,
\pi^+\pi^-,$ etc.  provide the main method used in the discovery of
new states from 27 years ago till now.  Four new bottomonium states
$h_b(1P), h_b(2P), Z_{b1}^+, Z_{b2}^+$ were observed at Belle, where
the $\Upsilon(5S)$ behaves like the particle $Y(4260)$ in the charm
sector.  Observations of the $h_b(1,2P)\to\eta_b(1S)\gamma$ and
$h_b(2P)\to\eta_b(2S)\gamma$ were made.  The particle $\eta_b(2S)$ was
observed for the first time and the hyperfine splitting $\Delta
M_{HF}(2S)$ at Belle agrees with theory.  S.Dobbs \textit{et al.}
obtained a higher value $\Delta M_{HF}(2S)$ from CLEO data and two
results are in clear disagreement.  For the other transitions of the
$\Upsilon(5S)$, observations of $\Upsilon(5S)\to\Upsilon(1,2S)\eta$
and $\Upsilon(5S)\to\Upsilon(1D)\pi^+\pi^-$ channels were made.  For
the $\eta$ transition of other $\Upsilon$ particles, The new
measurement of $\Upsilon(2S)\to\Upsilon(1S)\eta$ branching fraction is
still smaller than theory prediction.  Finally, In the talk covered by
Andre Chisholm, ATLAS and D0 has observed the $\chi_b(3P)$ state from
the transition $\chi_b(3P)\to \Upsilon(1,2S)\gamma$.  In future, we
expect more Belle results, and new results from the Hadron colliders.

\begin{acknowledgments}
The author Jin Li acknowledges support from
the BK21 and WCU program of the Ministry Education Science and
Technology, National Research Foundation of Korea Grant No.
2010-0021174, 2011-0029457,
2012-0008143, NRF-2012R1A1A2008330.

\end{acknowledgments}


\begin{thebibliography}{99}   
\bibitem{Herb:1977ek} 
  S.~W.~Herb, D.~C.~Hom, L.~M.~Lederman, J.~C.~Sens, H.~D.~Snyder, J.~K.~Yoh, J.~A.~Appel and B.~C.~Brown {\it et al.},
  Phys.\ Rev.\ Lett.\  {\bf 39}, 252 (1977).

\bibitem{Nernst:1985nn} 
  R.~Nernst {\it et al.}  [Crystal Ball Collaboration],
  Phys.\ Rev.\ Lett.\  {\bf 54}, 2195 (1985).

\bibitem{Aubert:2008ba} 
  B.~Aubert {\it et al.}  [BABAR Collaboration],
  Phys.\ Rev.\ Lett.\  {\bf 101}, 071801 (2008)
  [Erratum-ibid.\  {\bf 102}, 029901 (2009)]
  [arXiv:0807.1086 [hep-ex]].

\bibitem{Bonvicini:2004yj} 
  G.~Bonvicini {\it et al.}  [CLEO Collaboration],
  Phys.\ Rev.\ D {\bf 70}, 032001 (2004)
  [hep-ex/0404021].

\bibitem{Abe:2007tk} 
  K.~F.~Chen {\it et al.}  [Belle Collaboration],
  Phys.\ Rev.\ Lett.\  {\bf 100}, 112001 (2008)
  [arXiv:0710.2577 [hep-ex]].

\bibitem{Chen:2008xia} 
  K.~-F.~Chen {\it et al.}  [Belle Collaboration],
  Phys.\ Rev.\ D {\bf 82}, 091106 (2010)
  [arXiv:0810.3829 [hep-ex]].

\bibitem{CLEO:2011aa} 
  T.~K.~Pedlar {\it et al.}  [CLEO Collaboration],
  Phys.\ Rev.\ Lett.\  {\bf 107}, 041803 (2011)
  [arXiv:1104.2025 [hep-ex]].

\bibitem{Mo:2006ss} 
  X.~H.~Mo, G.~Li, C.~Z.~Yuan, K.~L.~He, H.~M.~Hu, J.~H.~Hu, P.~Wang and Z.~Y.~Wang,
  Phys.\ Lett.\ B {\bf 640}, 182 (2006)
  [hep-ex/0603024].

\bibitem{Adachi:2011ji} 
  I.~Adachi {\it et al.}  [Belle Collaboration],
  Phys.\ Rev.\ Lett.\  {\bf 108}, 032001 (2012)
  [arXiv:1103.3419 [hep-ex]].

\bibitem{Belle:2011aa} 
  A.~Bondar {\it et al.}  [Belle Collaboration],
  Phys.\ Rev.\ Lett.\  {\bf 108}, 122001 (2012)
  [arXiv:1110.2251 [hep-ex]].

\bibitem{Godfrey:2002rp} 
  S.~Godfrey and J.~L.~Rosner,
  Phys.\ Rev.\ D {\bf 66}, 014012 (2002)
  [hep-ph/0205255].

\bibitem{Mizuk:2012pb} 
  R.~Mizuk {\it et al.}  [Belle Collaboration],
  arXiv:1205.6351 [hep-ex].

\bibitem{Meinel:2010pv} 
  S.~Meinel,
  Phys.\ Rev.\ D {\bf 82}, 114502 (2010)
  [arXiv:1007.3966 [hep-lat]].

\bibitem{Dobbs:2012zn} 
  S.~Dobbs, Z.~Metreveli, K.~K.~Seth, A.~Tomaradze and T.~Xiao,
  Phys.\ Rev.\ Lett.\  {\bf 109}, 082001 (2012)
  [arXiv:1204.4205 [hep-ex]].


\bibitem{Voloshin:2007dx} 
  M.~B.~Voloshin,
  Prog.\ Part.\ Nucl.\ Phys.\  {\bf 61}, 455 (2008)
 [arXiv:0711.4556 [hep-ph]];
  Y.~-P.~Kuang,
  Front.\ Phys.\ China {\bf 1}, 19 (2006)
  [hep-ph/0601044].

\bibitem{He:2008xk} 
  Q.~He {\it et al.}  [CLEO Collaboration],
  Phys.\ Rev.\ Lett.\  {\bf 101}, 192001 (2008)
  [arXiv:0806.3027 [hep-ex]];
  J.~P.~Lees {\it et al.}  [BABAR Collaboration],
  Phys.\ Rev.\ D {\bf 84}, 092003 (2011)
  [arXiv:1108.5874 [hep-ex]].

\bibitem{Aubert:2008az} 
  B.~Aubert {\it et al.}  [BABAR Collaboration],
  Phys.\ Rev.\ D {\bf 78}, 112002 (2008)
  [arXiv:0807.2014 [hep-ex]].

\end{thebibliography}
\end{document}